\documentclass[a4paper,fleqn]{cas-sc}

\usepackage[numbers]{natbib}
\usepackage{amssymb}
\usepackage{amsmath}
\usepackage{graphicx}
\usepackage{booktabs}
\usepackage{xcolor}
\usepackage{float}
\usepackage{placeins}
\usepackage{caption}

\begin{document}
\let\WriteBookmarks\relax
\setcounter{topnumber}{2}
\renewcommand{\topfraction}{0.85}
\renewcommand{\textfraction}{0.1}
\renewcommand{\floatpagefraction}{0.8}

\shorttitle{RA-PINN for Irregular Interfaces and Multi-Peak Transport Fields}

\shortauthors{B.Zhou et al.}

\title [mode=title]{A Residual-Attention Physics-Informed Neural Network for Irregular Interfaces and Multi-Peak Transport Fields}

\author[1]{Baitong Zhou}[orcid=0009-0008-9546-8437]
\credit{Calculation, data analyzing and manuscript writing}\fnref{co-first}

\author[1]{Ze Tao}[orcid=0009-0004-0202-3641]
\credit{Calculation, data analyzing and manuscript writing}\fnref{co-first}

\affiliation[1]{organization={Nanophotonics and Biophotonics Key Laboratory of Jilin Province, School of Physics, Changchun University of Science and Technology},
                city={Changchun},
                postcode={130022},
                country={P.R. China}}

\affiliation[2]{organization={School of Chemistry and Environmental Engineering, Changchun University of Science and Technology},
                city={Changchun},
                postcode={130022},
                country={P.R. China}}

\author[1]{Fujun Liu}[orcid=0000-0002-8573-450X]
\credit{Review and Editing}
\cormark[1]

\author[1,2]{Xuan Fang}[orcid=0000-0000-0000-0000]
\credit{Review and Editing}
\cormark[1]

\fntext[co-first]{The authors contribute equally to this work.}
\ead{fjliu@cust.edu.cn}
\ead{fangx@cust.edu.cn}
\cortext[1]{Corresponding author}

\begin{abstract}
In complex engineering systems such as electro-thermal-fluid coupling, rapid and accurate prediction of multi-physics fields is essential for advanced applications like digital twins and real-time condition monitoring. Traditional numerical methods often suffer from high computational latency, whereas standard Physics-Informed Neural Networks (PINNs) frequently fail to capture critical local features, such as irregular interfaces, localized high-gradient regions, and multi-peak transport structures. To address these limitations and provide high-fidelity intelligent predictions for engineering decision-making, this paper proposes a Residual-Attention Physics-Informed Neural Network (RA-PINN) as a powerful surrogate modeling engine. The proposed method incorporates residual learning and attention enhancement into the network backbone to improve the representation of oblique transition structures, narrow charge layers, and distributed hotspots while strictly preserving global field consistency. To evaluate its effectiveness as an intelligent prediction framework, three representative benchmark cases are constructed, including an oblique asymmetric interface, a bipolar high-gradient charge layer, and a multi-peak Gaussian charge migration field. Under unified training settings, the proposed RA-PINN is systematically compared with a standard pure PINN and an LSTM-PINN in terms of average error, local maximum error, structural similarity, and convergence behavior. The results show that RA-PINN consistently achieves the best overall performance across all benchmark cases, demonstrating its tremendous potential as a highly reliable core inference engine for the condition monitoring and digital twin modeling of complex multi-physics engineering systems.
\end{abstract}


\begin{highlights}
\item A residual-attention PINN is developed for complex coupled multi-physics fields.
\item The proposed framework targets irregular interfaces, steep local gradients, and multi-peak transport structures.
\item RA-PINN improves local structural fidelity while preserving global field consistency.
\item RA-PINN consistently outperforms pure PINN and LSTM-PINN under unified training settings.
\end{highlights}


\maketitle
\section{Introduction}

Complex coupled multi-physics fields are widely encountered in engineering applications such as electro-thermal-fluid systems, microscale transport, electrohydrodynamics, and charge-regulated flow control. In these advanced engineering domains, the rapid and accurate prediction of physical states is crucial for deploying intelligent frameworks like digital twins and real-time condition monitoring systems. However, velocity, pressure, temperature, electric potential, and charge distribution are strongly coupled and often exhibit irregular interfaces, localized high-gradient regions, and multi-peak transport patterns. These local complex structures are of considerable engineering importance because interface position, hotspot intensity, and steep transition behavior are directly related to system performance, operational stability, and design reliability. Accurately reconstructing such coupled fields remains a significant challenge, especially when the target solution contains oblique transition layers, narrow bipolar high-gradient structures, or multiple distributed hotspots. Consequently, developing highly reliable intelligent modeling methods for these complex local structures is essential for empowering engineering-oriented prediction, automated analysis, and informed decision-making.

To address these challenges, a large number of numerical and learning-based methods have been developed in recent years. Traditional mesh-based numerical approaches have achieved considerable success in solving coupled-field problems, but they suffer from high computational latency and often require fine meshes, local refinement, or case-dependent numerical treatment when sharp interfaces and strongly localized structures are present. These computational bottlenecks make traditional methods poorly suited for real-time engineering applications. More recently, Physics-Informed Neural Networks (PINNs) have attracted significant attention as powerful surrogate models because they can integrate governing equations, boundary conditions, and observational information into a unified learning framework \cite{raissi2019pinn,karniadakis2021piml}. Existing studies have demonstrated the potential of PINNs as intelligent inference engines in fluid mechanics, heat transfer, and multi-physics modeling \cite{cai2021heattransfer,jagtap2020xpinns,yu2022gpinn,chen2025pfpinns}. Very recent studies have further extended physics-informed modeling toward data-efficient surrogate construction and engineering expert systems, especially in the context of Expert Systems with Applications \cite{jeong2024data,wu2024reviewpiml,zhang2026multienergy,wang2026gapinn}. In parallel, PINN-based formulations have also been advanced for difficult dynamic interfaces, physics-only training paradigms, and nonlinear transient heat conduction, which further highlight the importance of strengthening local-structure representation in scientific machine learning \cite{xing2025modeling,tao2025lnn,tao2025analytical}. Nevertheless, standard PINNs still suffer from optimization difficulty, spectral bias, and limited local structural fidelity when facing irregular interfaces, steep local gradients, and nonuniform multi-peak distributions \cite{wang2021gradient,wang2022ntk,sarma2024ipinns,tseng2023cusp}. In many practical scenarios, ordinary deep learning models can capture the smooth global field trend reasonably well but inherently ignore or underestimate the most critical local structures, such as hotspots and charge accumulation zones. This indicates that although existing studies have made substantial progress, the accurate representation of difficult local patterns in coupled engineering fields remains insufficient for building trustworthy expert systems.

In recent years, improved neural architectures have provided a promising direction for enhancing the performance and reliability of PINNs in intelligent prediction tasks. In particular, residual learning has been shown to improve optimization stability and information propagation in deep networks \cite{he2016resnet}, while attention mechanisms can significantly enhance the sensitivity of the model to informative features and important local responses \cite{vaswani2017attention}. Recurrent architectures such as LSTM have also been widely adopted for nonlinear feature propagation and long-range dependency modeling \cite{hochreiter1997lstm,wang2025pirn}. These developments suggest that architecture enhancement is a highly effective way to elevate the representational capability of PINNs from basic mathematical solvers to advanced surrogate modeling engines. Among these advancements, combining residual and attention mechanisms is particularly promising for coupled-field problems containing both smooth global backgrounds and localized difficult regions, and this trend is also consistent with recent ESWA studies emphasizing physics-informed and attention-aware surrogate learning for complex engineering tasks \cite{jeong2024data,wang2026gapinn}. Such a combined design is expected to preserve large-scale physical consistency while simultaneously strengthening the high-fidelity modeling of sharp transitions, high-gradient layers, and hotspot regions. Therefore, a residual-attention-enhanced PINN framework offers a robust intelligent solution to the current limitations of standard PINNs.

Motivated by this vision, this paper develops a Residual-Attention Physics-Informed Neural Network (RA-PINN) specifically designed to act as a core inference engine for irregular interfaces and multi-peak transport fields. The proposed intelligent method improves the representation of locally difficult structures while strictly maintaining the physics-constrained training paradigm of PINNs, thereby providing high-fidelity predictions for engineering decision-making. To rigorously evaluate its effectiveness as a condition monitoring surrogate, three representative benchmark cases are considered, namely an oblique asymmetric interface, a bipolar high-gradient charge layer, and a multi-peak Gaussian charge migration field. These three cases map directly to three typical forms of engineering difficulty: irregular transition geometry, narrow steep-gradient structures, and multiple distributed hotspots. Under unified training settings, the proposed RA-PINN is systematically compared with a standard pure PINN and an LSTM-PINN. The results show that RA-PINN consistently achieves the best overall performance in all three cases, exhibiting a significantly more accurate recovery of oblique interfaces, localized high-gradient layers, and multi-peak spatial distributions. These findings demonstrate that the proposed residual-attention mechanism provides an effective, high-precision, and engineering-relevant enhancement for the intelligent modeling of complex coupled multi-physics fields.

The paper is organized as follows. Section 2 introduces the governing equations and benchmark formulation. Section 3 presents the proposed residual-attention PINN methodology and its intelligent architecture. Section 4 describes the experimental setup used to evaluate the models. Section 5 reports the numerical results and provides a comparative analysis of the prediction capabilities. Section 6 discusses the engineering implications of the proposed method for advanced surrogate modeling. Finally, Section 7 concludes the paper.

\section{Governing Equations and Problem Description}

\subsection{Governing equations}

In this work, a coupled multi-physics field system is considered, involving velocity components, pressure, temperature, and electric potential. Let the computational domain be denoted by $\Omega \subset \mathbb{R}^2$, with spatial coordinates $\mathbf{x}=(x,y)$. The target solution is written as
\begin{equation}
\mathbf{U}(x,y) = \left[ u(x,y),\, v(x,y),\, p(x,y),\, T(x,y),\, \phi(x,y) \right],
\end{equation}
where $u$ and $v$ are the velocity components, $p$ is the pressure, $T$ is the temperature, and $\phi$ is the electric potential.

The coupled governing equations are composed of the continuity equation, momentum equations, energy equation, and electric potential equation. In a general form, they can be written as
\begin{equation}
\nabla \cdot \mathbf{u} = 0,
\end{equation}
\begin{equation}
\mathbf{u}\cdot\nabla u + \frac{\partial p}{\partial x} - \nu \nabla^2 u - f_x(\phi,T) = s_u(x,y),
\end{equation}
\begin{equation}
\mathbf{u}\cdot\nabla v + \frac{\partial p}{\partial y} - \nu \nabla^2 v - f_y(\phi,T) = s_v(x,y),
\end{equation}
\begin{equation}
\mathbf{u}\cdot\nabla T - \alpha \nabla^2 T - q_T(\phi) = s_T(x,y),
\end{equation}
\begin{equation}
\nabla^2 \phi - \lambda_\phi \phi - q(x,y) = s_\phi(x,y),
\end{equation}
where $\mathbf{u}=(u,v)$ is the velocity vector, $\nu$ is the viscous coefficient, $\alpha$ is the thermal diffusivity, $\lambda_\phi$ is the electric potential reaction coefficient, $q(x,y)$ denotes the prescribed charge-related source distribution, and $s_u$, $s_v$, $s_T$, and $s_\phi$ are source terms determined from the prescribed reference solutions. The terms $f_x(\phi,T)$, $f_y(\phi,T)$, and $q_T(\phi)$ represent the coupled electric and thermal driving effects in the momentum and energy equations.

\subsection{Benchmark formulation with known reference solutions}

To systematically evaluate the representational capability of different PINN architectures in complex coupled fields, this study adopts a benchmark framework with known analytic reference solutions. Instead of relying on experimental measurements or high-fidelity finite element simulations, analytic target fields are first constructed, and the corresponding source terms are then derived so that the prescribed fields exactly satisfy the governing equations. This rigorous approach is intentionally chosen to completely eliminate the unpredictable interference of experimental data noise and numerical discretization errors. By isolating these external factors, the framework ensures an absolutely fair and strict quantitative assessment of how different network architectures, namely the pure PINN, LSTM-PINN, and the proposed RA-PINN, handle critical local features.

Let the exact solution be denoted by
$$U^{\ast}(x,y)=[u^{\ast}(x,y),v^{\ast}(x,y),p^{\ast}(x,y),T^{\ast}(x,y),\phi^{\ast}(x,y)].$$
By substituting $U^{\ast}$ into the governing equations, the corresponding source terms are obtained as
$$s_u=\mathcal{R}_u(U^{\ast}),\quad s_v=\mathcal{R}_v(U^{\ast}),\quad s_T=\mathcal{R}_T(U^{\ast}),\quad s_\phi=\mathcal{R}_\phi(U^{\ast}),$$
where $\mathcal{R}_u$, $\mathcal{R}_v$, $\mathcal{R}_T$, and $\mathcal{R}_\phi$ denote the corresponding differential operators of the momentum, energy, and electric potential equations.

This benchmark construction provides two primary advantages for evaluating intelligent prediction models. First, the exact field values are available at any spatial location, making it possible to quantitatively evaluate prediction errors across the entire domain using metrics such as RMSE, MAE, and maximum absolute error. Second, diverse physical patterns can be directly embedded into the analytic fields to construct representative benchmark cases that topologically map to genuine engineering challenges. In this paper, three representative benchmark cases are considered, namely Case 1, Case 2, and Case 3. These cases are carefully selected because they perfectly mirror the most typical sources of difficulty encountered in engineering-oriented coupled-field modeling.

Case 1 represents an oblique asymmetric transition interface. The corresponding field construction contains inclined transition structures, making it highly suitable for testing the ability of the model to capture non-axisymmetric interfaces and directional discontinuity-like regions often found in complex fluid boundaries. Building upon the challenge of spatial transitions, Case 2 represents a bipolar high-gradient charge layer. In this scenario, the prescribed field exhibits a rapid positive-negative variation within an extremely narrow region, providing a demanding benchmark for evaluating local gradient reconstruction and transition-layer resolution critical to electro-thermal applications. Finally, Case 3 represents a multi-peak Gaussian charge migration field. The charge-related source distribution is composed of multiple spatial peaks combined with an additional oscillatory perturbation, resulting in a highly nonuniform coupled field. This configuration rigorously tests the ability of the intelligent model to simultaneously reconstruct multiple distributed hotspots and complex transport paths without artificial smoothing.

Together, these three cases comprehensively cover irregular interfaces, strong local gradients, and multi-peak nonuniform distributions. Consequently, they form a compact yet highly representative benchmark set that directly reflects real-world engineering topologies, establishing a solid foundation for evaluating the practical applicability and structural fidelity of the proposed RA-PINN framework against baseline architectures.

\subsection{Boundary conditions, supervised data, and learning objective}

For all benchmark cases, the exact reference solutions are used to generate boundary data and supervised reference samples. Specifically, boundary values are directly extracted from the exact fields,
\begin{equation}
\mathbf{U}(x,y)\big|_{\partial\Omega} = \mathbf{U}^{\ast}(x,y)\big|_{\partial\Omega},
\end{equation}
where $\partial\Omega$ denotes the domain boundary.

In addition to boundary constraints, a set of supervised interior samples is constructed from the exact solution to assist training and improve convergence stability. Therefore, the training process combines both physics constraints and data constraints. This hybrid setting is particularly suitable for the present benchmark study, because it allows the comparison to focus on the representational capability of different network architectures under the same governing equations and the same supervision conditions.

Given a neural network approximation $\hat{\mathbf{U}}(x,y)$, the objective is to minimize the discrepancy between the predicted fields and the exact fields while satisfying the governing equations and boundary conditions. Accordingly, the learning objective consists of three parts:
\begin{equation}
\mathcal{L} = \mathcal{L}_{\mathrm{data}} + \mathcal{L}_{\mathrm{bc}} + \mathcal{L}_{\mathrm{pde}},
\end{equation}
where $\mathcal{L}_{\mathrm{data}}$ is the supervised data loss, $\mathcal{L}_{\mathrm{bc}}$ is the boundary-condition loss, and $\mathcal{L}_{\mathrm{pde}}$ is the PDE residual loss. The detailed network construction and loss implementation are introduced in the next section.

The purpose of this study is not only to obtain accurate field prediction, but also to compare how different PINN architectures perform when facing irregular interfaces, localized high-gradient layers, and multi-peak transport structures. Therefore, under the same physical formulation and the same training settings, the comparison among pure PINN, LSTM-PINN, and the proposed residual-attention PINN can directly reveal the effect of network architecture on complex coupled-field modeling.

\section{Residual-Attention PINN Method}

\subsection{PINN formulation and compared models}

Physics-Informed Neural Networks (PINNs) approximate the target field variables by a neural network and enforce the governing equations through automatic differentiation \cite{raissi2019pinn,karniadakis2021piml}. For the present problem, the input of the network is the spatial coordinate $\mathbf{x}=(x,y)$, and the output is the coupled field vector
\begin{equation}
\hat{\mathbf{U}}(x,y)=\left[\hat{u}(x,y),\,\hat{v}(x,y),\,\hat{p}(x,y),\,\hat{T}(x,y),\,\hat{\phi}(x,y)\right].
\end{equation}

Given the predicted fields, spatial derivatives are obtained by automatic differentiation, and the corresponding PDE residuals can be written as
\begin{equation}
\mathbf{R}(x,y)=\left[R_{\mathrm{cont}},\,R_u,\,R_v,\,R_T,\,R_{\phi}\right].
\end{equation}
The standard PINN training objective is constructed by minimizing the discrepancy on supervised interior samples, the mismatch on boundary samples, and the residuals of the governing equations. Therefore, the loss function takes the form
\begin{equation}
\mathcal{L}=\mathcal{L}_{\mathrm{data}}+\mathcal{L}_{\mathrm{bc}}+\mathcal{L}_{\mathrm{pde}},
\end{equation}
where
\begin{equation}
\mathcal{L}_{\mathrm{data}}=\frac{1}{N_d}\sum_{i=1}^{N_d}\left\|\hat{\mathbf{U}}(\mathbf{x}_i^d)-\mathbf{U}^{\ast}(\mathbf{x}_i^d)\right\|_2^2,
\end{equation}
\begin{equation}
\mathcal{L}_{\mathrm{bc}}=\frac{1}{N_b}\sum_{i=1}^{N_b}\left\|\hat{\mathbf{U}}(\mathbf{x}_i^b)-\mathbf{U}^{\ast}(\mathbf{x}_i^b)\right\|_2^2,
\end{equation}
and
\begin{equation}
\mathcal{L}_{\mathrm{pde}}=\frac{1}{N_r}\sum_{i=1}^{N_r}\left(
|R_{\mathrm{cont}}|^2+|R_u|^2+|R_v|^2+|R_T|^2+|R_{\phi}|^2
\right).
\end{equation}
Here, $N_d$, $N_b$, and $N_r$ denote the numbers of interior supervised samples, boundary samples, and residual points, respectively.

Although this formulation is conceptually straightforward, standard PINNs often struggle when the target fields contain oblique interfaces, high-gradient local layers, or multi-peak nonuniform structures \cite{wang2021gradient,wang2022ntk,tancik2020fourier,sitzmann2020siren}. In such situations, the network may fit the smooth background reasonably well but still fail to reconstruct local critical regions with sufficient accuracy. This motivates the introduction of improved network architectures for complex engineering-oriented coupled-field problems.

To provide a fair comparison, three network architectures are considered in this study: pure PINN, LSTM-PINN, and the proposed Residual-Attention PINN (RA-PINN). All three models share the same physical formulation, the same loss definition, and the same training data configuration. Their difference lies only in the network backbone.

The pure PINN adopts a standard fully connected multilayer perceptron (MLP) to approximate the target field vector. This model serves as the baseline architecture and represents the conventional PINN framework without additional structure enhancement.

The LSTM-PINN introduces an LSTM-based feature propagation mechanism into the network backbone. Compared with the pure PINN, this model is expected to improve feature interaction and sequence-like hidden-state propagation. In the present study, it is used as a stronger comparison baseline beyond the standard MLP-based PINN, motivated by the long short-term memory architecture originally developed for sequential dependency modeling \cite{hochreiter1997lstm} and is also related to recent physics-informed recurrent modeling explorations \cite{wang2025pirn}.

The proposed RA-PINN incorporates both residual connections and attention enhancement into the backbone. The residual structure is introduced to stabilize deep feature propagation and preserve coarse-scale background information \cite{he2016resnet}, whereas the attention mechanism is designed to strengthen the response of the network to locally important regions such as irregular transition zones, steep-gradient layers, and multiple hotspots \cite{vaswani2017attention}.

For all three models, the hidden feature dimension is fixed as 128. The pure PINN uses a six-layer fully connected multilayer perceptron with Tanh activation functions. The LSTM-PINN employs an input projection layer followed by two stacked single-layer LSTM modules and a nonlinear output head. The proposed RA-PINN is constructed with six residual-attention blocks under the same hidden width, followed by a nonlinear output head.

\subsection{Residual-attention network architecture}
\begin{figure}[H]
    \centering
    \includegraphics[width=0.96\textwidth]{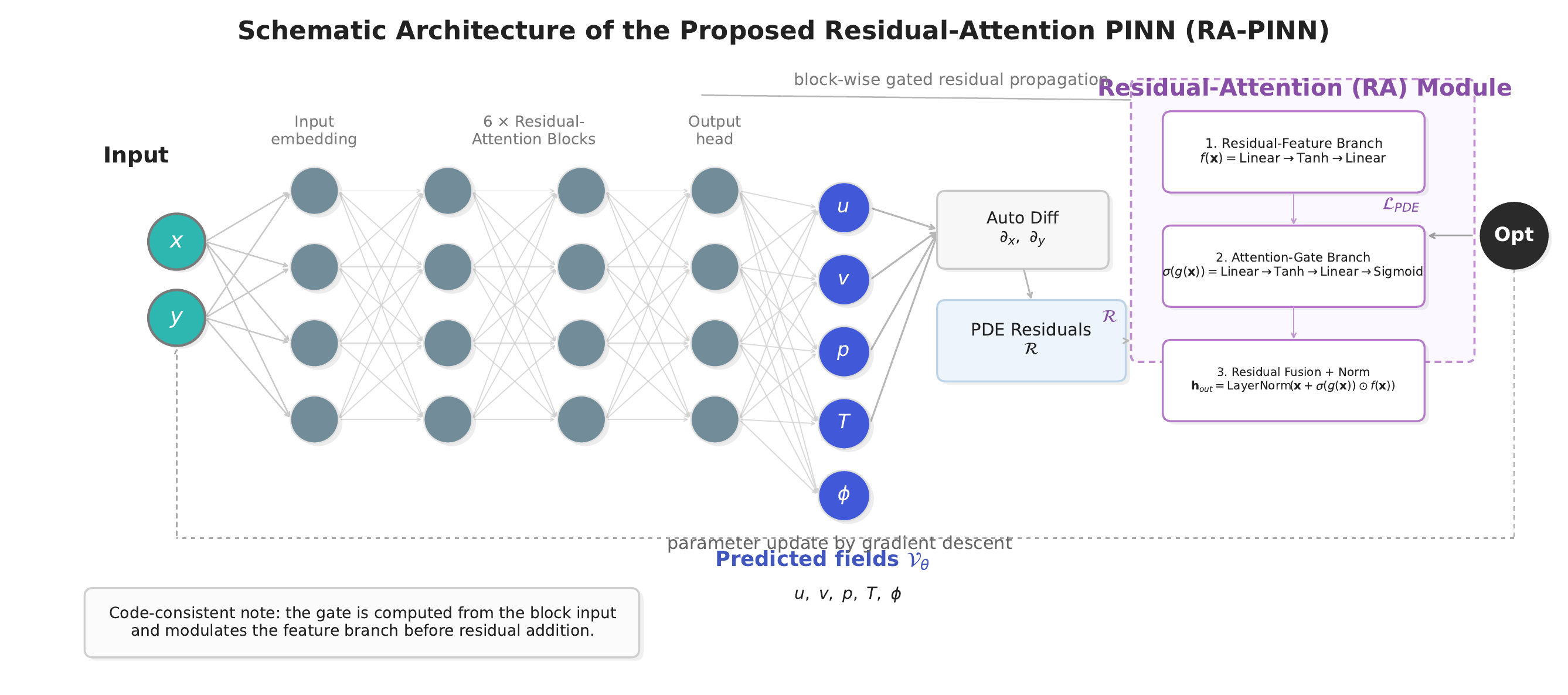}
    \caption{Schematic architecture of the proposed residual-attention physics-informed neural network (RA-PINN). The network takes the spatial coordinates $(x,y)$ as input and predicts five coupled physical fields $(u,v,p,T,\phi)$. Automatic differentiation is used to compute the required spatial derivatives and construct the PDE residuals for physics-informed training. As shown in the residual-attention module, the hidden feature is processed through a residual-feature branch and an attention-gate branch, where the gate is generated from the block input and adaptively modulates the feature branch before residual addition, followed by layer normalization. This design improves the representation of irregular interfaces, localized high-gradient regions, and multi-peak field structures while preserving global feature propagation.}
    \label{fig:ra_pinn_architecture}
\end{figure}
\FloatBarrier

Fig.\ref{fig:ra_pinn_architecture} shows the overall architecture of the proposed RA-PINN. The model takes the spatial coordinates $(x,y)$ as input and predicts five coupled physical fields, namely $u$, $v$, $p$, $T$, and $\phi$. These predicted fields are further differentiated through automatic differentiation to construct the PDE residuals required by the physics-informed learning framework. In the backbone network, multiple residual-attention blocks are stacked to improve the representation of difficult local structures. As illustrated in the right part of Fig.\ref{fig:ra_pinn_architecture}, each residual-attention module contains a residual-feature branch and an attention-gate branch. The gate is generated from the block input and is used to adaptively modulate the feature branch before residual addition, followed by layer normalization. Through this design, the proposed network can preserve stable background information while enhancing the sensitivity to irregular interfaces, localized high-gradient regions, and multi-peak spatial patterns.

Let $\mathbf{h}^{(l)}$ denote the hidden feature at layer $l$. In the residual-attention backbone, the nonlinear feature branch of a block can be written as
\begin{equation}
\mathbf{f}^{(l)}=\mathcal{F}\left(\mathbf{h}^{(l)};\Theta_f^{(l)}\right),
\end{equation}
where $\mathcal{F}(\cdot)$ denotes the residual-feature transformation parameterized by $\Theta_f^{(l)}$. In the present implementation, this branch is realized by two fully connected transformations with nonlinear activation. Residual learning helps alleviate optimization degradation and facilitates information propagation across layers \cite{he2016resnet}.

To enhance local feature sensitivity, an attention-gate branch is introduced in parallel. The gate is generated directly from the block input and can be expressed as
\begin{equation}
\mathbf{g}^{(l)}=\sigma\!\left(\mathcal{G}\left(\mathbf{h}^{(l)};\Theta_g^{(l)}\right)\right),
\end{equation}
where $\mathcal{G}(\cdot)$ denotes the gate-generating transformation, $\Theta_g^{(l)}$ is the corresponding parameter set, and $\sigma(\cdot)$ denotes the sigmoid activation used to produce bounded channel-wise modulation coefficients. The gated residual interaction is then written as
\begin{equation}
\mathbf{h}^{(l+1)}=\mathrm{LayerNorm}\!\left(\mathbf{h}^{(l)}+\mathbf{g}^{(l)}\odot \mathbf{f}^{(l)}\right),
\end{equation}
where $\odot$ denotes element-wise multiplication. In this form, the attention gate does not directly reweight the hidden feature itself; instead, it adaptively modulates the residual-feature branch before residual addition. This mechanism enables the network to place more emphasis on regions or patterns that are more difficult to model, such as sharp local transitions and multi-peak distributions \cite{vaswani2017attention}.

According to the current implementation, the input coordinates are first projected into the hidden feature space, then propagated through six stacked residual-attention blocks, and finally mapped to the five target variables through the output head. Therefore, the proposed RA-PINN can be interpreted as a gated residual feature propagation network embedded into the standard physics-informed learning framework. Compared with a plain multilayer perceptron, this design is intended to preserve global structural consistency while improving the representation of localized difficult regions.

This combination is particularly suitable for the present benchmark set, because all selected cases contain either irregular interfaces, localized high-gradient regions, or multiple spatial peaks. The three benchmark cases considered in this study correspond to three representative forms of local difficulty. Case 1 contains an oblique asymmetric interface, which requires the model to distinguish directional transition patterns in a non-axisymmetric setting. Case 2 contains a bipolar high-gradient charge layer, which tests the ability of the network to reconstruct narrow regions with steep local variation. Case 3 contains a multi-peak Gaussian migration field, which requires the model to preserve multiple hotspot structures simultaneously.

For these problems, a single plain MLP may tend to smooth out local extreme features during optimization. Residual connections help maintain stable propagation of feature information across layers, which is useful for preserving the global field structure. However, global stability alone is insufficient for highly localized patterns. The gated attention branch compensates for this limitation by increasing the sensitivity of the network to difficult local regions through adaptive feature modulation. Therefore, the combination of residual learning and gated attention weighting is expected to be more effective than using a plain MLP when the target solution contains irregular interfaces, concentrated gradients, or multiple peaks.

\subsection{Training strategy}

All models are trained under the same optimization framework in order to ensure a fair comparison. The training samples consist of three parts: interior supervised points generated from the reference solution, boundary points extracted from the exact boundary conditions, and residual points used to enforce the governing equations. The total loss is defined by
\begin{equation}
\mathcal{L}_{\mathrm{total}}=
\mathcal{L}_{\mathrm{data}}+
\mathcal{L}_{\mathrm{bc}}+
\mathcal{L}_{\mathrm{pde}}.
\end{equation}

In practice, the PDE residual loss can also be written as the sum of individual residual components,
\begin{equation}
\mathcal{L}_{\mathrm{pde}}=
\mathcal{L}_{\mathrm{cont}}+
\mathcal{L}_{u}+
\mathcal{L}_{v}+
\mathcal{L}_{T}+
\mathcal{L}_{\phi},
\end{equation}
where the five terms correspond to the continuity, momentum, energy, and electric potential equations. This decomposition is useful for monitoring the training process and identifying whether a model has difficulty in a specific physical component.

The network parameters are optimized using the same optimizer and the same training schedule for all three architectures. Consequently, the comparison focuses on the effect of the network backbone rather than differences in training hyperparameters. The evaluation is conducted by comparing the predicted fields with the exact reference solutions in terms of average error, local maximum error, structural similarity, and convergence behavior.

The overall workflow of the proposed method can be summarized as follows:
\begin{enumerate}
\item Construct the benchmark-specific reference solution and derive the corresponding source terms;
\item Generate interior supervised samples, boundary samples, and residual collocation points;
\item Feed the spatial coordinates into the neural network and predict the coupled field variables;
\item Compute spatial derivatives by automatic differentiation and evaluate the PDE residuals;
\item Optimize the network parameters by minimizing the combined data, boundary, and PDE losses;
\item Compare the predicted fields of different architectures against the exact reference fields.
\end{enumerate}

This workflow is shared by the pure PINN, the LSTM-PINN, and the proposed RA-PINN. Therefore, any difference in the final prediction performance can be directly attributed to the representational capability of the network architecture itself. In the next section, the three benchmark cases and the corresponding experimental settings are introduced in detail.

\section{Benchmark Cases and Experimental Settings}

\subsection{Benchmark design}

To systematically evaluate the capability of different PINN architectures in engineering-oriented coupled-field problems, three benchmark cases are selected in this study, namely Case 1, Case 2, and Case 3. These cases are not chosen arbitrarily. Instead, they are designed to represent three typical sources of difficulty frequently encountered in complex multi-physics modeling:
\begin{enumerate}
\item irregular and oblique interface structures,
\item localized high-gradient transition layers,
\item multi-peak and nonuniform transport patterns.
\end{enumerate}

Accordingly, Case 1 is used to represent an oblique asymmetric transition interface, Case 2 is used to represent a bipolar high-gradient charge layer, and Case 3 is used to represent a multi-peak Gaussian migration field. These three cases form a compact benchmark set for evaluating whether a given network architecture can simultaneously maintain global field consistency and recover local complex structures.

Together, these benchmark cases provide a compact yet representative testbed for examining whether a PINN architecture can remain robust under different types of local structural difficulty encountered in engineering-oriented coupled-field modeling.

\subsection{Case 1: Oblique asymmetric interface}

Case 1 is constructed to represent an oblique asymmetric transition interface. In this case, the analytic field contains inclined transition structures that are not aligned with the Cartesian coordinate axes. As a result, the target solution presents a clear directional interface pattern with non-axisymmetric variation.

The main challenge of Case 1 lies in the reconstruction of oblique transition zones. Compared with an axis-aligned interface, an inclined interface is more difficult for a neural network to approximate accurately, because the spatial variation cannot be captured through simple one-directional feature adaptation. For standard PINNs, such oblique structures are often smoothed out during training, especially when the model tends to favor global average accuracy over local interface sharpness.

Therefore, Case 1 is particularly suitable for testing whether the proposed RA-PINN can improve the representation of directional transition layers and preserve local interface geometry more accurately than the baseline models.

\subsection{Case 2: Bipolar high-gradient charge layer}

Case 2 is designed to represent a bipolar high-gradient charge layer. The prescribed charge-related distribution exhibits rapid positive-negative variation within a narrow region, thereby forming a highly localized transition layer with steep gradients.

This case is intended to mimic engineering situations in which the field changes dramatically over a short spatial distance, such as narrow electric transition zones or highly localized charge accumulation regions. Such structures place strong demands on the ability of the network to resolve local gradients and maintain numerical stability during training.

For standard PINNs, these steep local transitions are often difficult to reconstruct because the model may prioritize smooth background fitting and underrepresent narrow high-gradient regions. Therefore, Case 2 provides a challenging benchmark for evaluating local gradient reconstruction capability and is particularly useful for assessing whether the residual-attention mechanism can enhance the sensitivity of the model to critical local layers.

\subsection{Case 3: Multi-peak Gaussian charge migration field}

Case 3 is used to represent a multi-peak Gaussian charge migration field. In this benchmark, the charge-related source distribution is composed of multiple Gaussian-like spatial peaks together with an additional oscillatory perturbation. Consequently, the target coupled field becomes highly nonuniform and contains several hotspot-like regions distributed over the computational domain.

The main challenge of this case lies in the simultaneous recovery of multiple local peaks. Unlike single-interface or single-gradient-layer problems, Case 3 requires the network to preserve several spatially separated but physically relevant regions at the same time. If the representational capability of the model is insufficient, some peaks may be smoothed out, shifted, or merged during training.

Therefore, Case 3 is adopted to evaluate whether the proposed RA-PINN can better reconstruct multiple hotspots and complex migration paths than the baseline pure PINN and LSTM-PINN models.

As can be seen, the selected cases cover three complementary forms of local complexity. Together, they allow a systematic investigation of whether a neural network architecture can remain robust under different types of engineering-oriented field difficulty.

\subsection{Experimental settings and evaluation metrics}

For each benchmark case, the exact reference solution is used to generate three types of training samples:
\begin{enumerate}
\item interior supervised points,
\item boundary-condition points,
\item PDE residual collocation points.
\end{enumerate}

The interior supervised points are sampled from the computational domain and paired with exact field values. These samples are used in the data loss term to guide the network toward the target solution. The boundary points are sampled along the domain boundary and are associated with exact boundary values extracted from the reference solution. These samples are used to enforce boundary consistency. The collocation points are used to evaluate the PDE residuals through automatic differentiation.

This hybrid physics-data setting is adopted to focus the comparison on the representational capability of the network architectures rather than on the availability of field information. Since the exact solutions are known, this setting enables a fair and controlled benchmark study for architecture comparison.

Three models are compared in this study:
\begin{itemize}
\item \textbf{pure PINN}: the standard fully connected PINN without additional feature enhancement,
\item \textbf{LSTM-PINN}: a PINN equipped with an LSTM-based feature propagation mechanism,
\item \textbf{RA-PINN}: the proposed residual-attention PINN.
\end{itemize}

To ensure fairness, all three models are trained under the same physical formulation, the same benchmark cases, and the same overall data-generation strategy. The comparison is therefore focused on the effect of the backbone architecture on complex coupled-field prediction.

To ensure a fair comparison, all models are trained under the same optimization framework. Specifically, the random seed is fixed as 20260308, and the computation is carried out on a GPU when CUDA is available, otherwise on a CPU. All models are trained for 50{,}000 optimization steps using the Adam optimizer with an initial learning rate of $8.0\times10^{-4}$ and a weight decay of $1.0\times10^{-10}$. The validation loss is evaluated every 200 training steps.

The computational domain is set to $\Omega=[-1,1]\times[-1,1]$ for all three benchmark cases. The supervised dataset is generated on a structured grid of $91\times91$ points, resulting in 8{,}281 supervised samples in total. These samples are randomly split into a training set and a validation set with a ratio of 0.7:0.3, yielding approximately 5{,}796 training samples and 2{,}485 validation samples. The PDE residual loss is evaluated using 16{,}000 interior collocation points sampled uniformly from the domain. Boundary constraints are imposed using 360 points on each boundary edge, resulting in 1{,}436 unique boundary samples after removing repeated corner points. For evaluation and visualization, a finer grid of $181\times181$ points is used.

The hidden feature dimension is fixed as 128 for all three architectures. The pure PINN adopts a six-layer fully connected multilayer perceptron with Tanh activation functions. The LSTM-PINN uses an input projection layer followed by two single-layer LSTM modules and a nonlinear output head. The proposed RA-PINN uses six residual-attention blocks built on the same hidden dimension, followed by a nonlinear output head. In addition, the PDE loss is introduced with a warm-up factor of 0.25 at the early stage of training to improve optimization stability. Under these unified settings, performance differences among the models can be attributed primarily to differences in network architecture rather than external hyperparameter tuning.

To quantitatively evaluate the prediction performance of each model, several standard metrics are used. Let $u_i^{\ast}$ and $\hat{u}_i$ denote the exact and predicted values at the $i$-th sample point. Then the root mean square error (RMSE) is defined as
\begin{equation}
\mathrm{RMSE}=\sqrt{\frac{1}{N}\sum_{i=1}^{N}\left(\hat{u}_i-u_i^{\ast}\right)^2},
\end{equation}
the mean absolute error (MAE) is defined as
\begin{equation}
\mathrm{MAE}=\frac{1}{N}\sum_{i=1}^{N}\left|\hat{u}_i-u_i^{\ast}\right|,
\end{equation}
and the maximum absolute error is defined as
\begin{equation}
\mathrm{MaxAbs}=\max_{1\le i\le N}\left|\hat{u}_i-u_i^{\ast}\right|.
\end{equation}

These metrics are computed for each physical variable and can also be averaged over all output fields for overall comparison. In addition to numerical errors, qualitative comparisons are also performed by visualizing the exact field, predicted field, and absolute error field. Furthermore, convergence behavior is examined through the evolution of total loss, PDE loss, and validation-related errors during training.

The purpose of this experimental design is not merely to test whether the models can fit the target fields. Instead, the main objective is to reveal how different architectures behave when the solution contains irregular interfaces, localized steep transitions, and multiple distributed hotspots. Since all three models are evaluated under the same benchmark framework and the same training conditions, the comparison can directly indicate whether the residual-attention mechanism provides a meaningful advantage in engineering-oriented coupled-field modeling.

The numerical results corresponding to these benchmark cases are presented and discussed in the next section.

For reproducibility, the main hyperparameter settings of the final residual-attention implementations for Cases 1, 2, and 3 are summarized in Appendix~A.

\section{Results and Discussion}

\subsection{Overall comparison}

This section presents the comparative results of the three models, namely pure PINN, LSTM-PINN, and the proposed RA-PINN, on the three benchmark cases introduced in Section 4. The comparison is conducted from multiple perspectives, including averaged prediction accuracy, local maximum error, structural consistency between the exact and predicted fields, convergence behavior during training, and computational cost.

Overall, the proposed RA-PINN achieves the best performance across the three benchmark cases and shows the most favorable balance between global accuracy and local structural fidelity. Compared with the pure PINN, the proposed model provides more accurate reconstruction of local transition regions, steep-gradient layers, and multiple hotspot structures. Compared with the LSTM-PINN, RA-PINN further improves the recovery of irregular spatial patterns and reduces error concentration in critical local regions. These results indicate that the residual-attention mechanism is beneficial for coupled-field modeling problems in which both global consistency and local structural accuracy are required.

In terms of overall quantitative performance, the three models can be ranked as
\begin{equation}
\mathrm{RA\text{-}PINN} \; > \; \mathrm{LSTM\text{-}PINN} \; > \; \mathrm{pure\ PINN}.
\end{equation}
This ranking is observed not only in averaged error metrics, but also in local maximum error control and visual similarity between the exact and predicted fields.

For Cases 1, 2, and 3, the averaged RMSE of RA-PINN is reduced to $5.83\times10^{-5}$, $8.01\times10^{-5}$, and $6.93\times10^{-5}$, respectively, compared with $2.32\times10^{-4}$, $3.85\times10^{-4}$, and $1.71\times10^{-4}$ for the pure PINN, and $1.68\times10^{-4}$, $2.12\times10^{-4}$, and $1.59\times10^{-4}$ for the LSTM-PINN. These results show that the proposed RA-PINN consistently provides the most accurate field prediction among the three architectures.

Table~\ref{tab:overall_comparison_case127} summarizes the averaged prediction errors of the three models on the three benchmark cases. The proposed RA-PINN consistently achieves the lowest averaged RMSE, MSE, MAE, L2 error, and maximum absolute error in all three cases. This consistency suggests that the residual-attention mechanism provides a meaningful improvement in coupled-field modeling rather than a case-specific advantage.

\begin{center}
\captionsetup{type=table}
\captionof{table}{Overall comparison of the three models on Cases 1, 2, and 3. The averaged values are computed over the five output fields.}
\label{tab:overall_comparison_case127}
\resizebox{\linewidth}{!}{%
\begin{tabular}{p{1.2cm} p{2.2cm} p{1.8cm} p{1.8cm} p{1.8cm} p{1.8cm} p{2.0cm}}
\toprule
Case & Model & Avg.\ RMSE & Avg.\ MSE & Avg.\ MAE & Avg.\ L2 error & Avg.\ MaxAbs \\
\midrule
\multirow{3}{*}{Case 1}
& pure PINN & $2.32\times10^{-4}$ & $5.53\times10^{-8}$ & $1.77\times10^{-4}$ & $3.87\times10^{-4}$ & $1.69\times10^{-3}$ \\
& LSTM-PINN & $1.68\times10^{-4}$ & $2.90\times10^{-8}$ & $1.24\times10^{-4}$ & $2.76\times10^{-4}$ & $1.29\times10^{-3}$ \\
& RA-PINN & $\mathbf{5.83\times10^{-5}}$ & $\mathbf{3.53\times10^{-9}}$ & $\mathbf{4.56\times10^{-5}}$ & $\mathbf{9.73\times10^{-5}}$ & $\mathbf{3.43\times10^{-4}}$ \\
\midrule
\multirow{3}{*}{Case 2}
& pure PINN & $3.85\times10^{-4}$ & $1.54\times10^{-7}$ & $2.86\times10^{-4}$ & $1.04\times10^{-3}$ & $2.45\times10^{-3}$ \\
& LSTM-PINN & $2.12\times10^{-4}$ & $4.61\times10^{-8}$ & $1.57\times10^{-4}$ & $5.62\times10^{-4}$ & $1.37\times10^{-3}$ \\
& RA-PINN & $\mathbf{8.01\times10^{-5}}$ & $\mathbf{6.24\times10^{-9}}$ & $\mathbf{6.14\times10^{-5}}$ & $\mathbf{2.07\times10^{-4}}$ & $\mathbf{5.56\times10^{-4}}$ \\
\midrule
\multirow{3}{*}{Case 3}
& pure PINN & $1.71\times10^{-4}$ & $2.97\times10^{-8}$ & $1.32\times10^{-4}$ & $6.84\times10^{-4}$ & $1.11\times10^{-3}$ \\
& LSTM-PINN & $1.59\times10^{-4}$ & $2.56\times10^{-8}$ & $1.19\times10^{-4}$ & $6.74\times10^{-4}$ & $9.77\times10^{-4}$ \\
& RA-PINN & $\mathbf{6.93\times10^{-5}}$ & $\mathbf{5.10\times10^{-9}}$ & $\mathbf{5.55\times10^{-5}}$ & $\mathbf{3.04\times10^{-4}}$ & $\mathbf{4.58\times10^{-4}}$ \\
\bottomrule
\end{tabular}%
}
\end{center}

Beyond averaged metrics such as RMSE and MAE, local maximum error and structural consistency are also important for evaluating prediction quality in coupled-field problems. In practical applications, a model with a small global average error may still be unreliable if it fails to accurately recover critical local structures such as interfaces, hotspots, or steep transition layers.

Across the three benchmark cases, the pure PINN generally produces larger local error concentrations near oblique interfaces, narrow gradient layers, and hotspot regions. The LSTM-PINN reduces these errors to some extent, but noticeable local mismatch remains in the most difficult regions of each case. In contrast, the proposed RA-PINN consistently exhibits smaller error concentration regions and better overlap between the predicted structures and the reference structures. This improvement can be interpreted from the network design: residual learning helps preserve stable feature propagation across layers, while the gated attention mechanism increases the sensitivity of the model to spatially critical regions.

The convergence histories of the three models further support this observation. For all benchmark cases, the training losses exhibit stable decreasing trends, but noticeable differences exist in the final convergence level and optimization smoothness. The pure PINN generally converges to the highest final loss level, whereas the LSTM-PINN reaches a lower loss level with improved stability. The proposed RA-PINN typically achieves the lowest or most stable convergence platform, indicating that the residual-attention enhancement is beneficial not only for final accuracy but also for the optimization process itself.

In addition to prediction accuracy, the computational cost of different architectures is also important for practical applications. Since all models are trained under the same optimization setting with a fixed training budget of 50,000 steps, the computational overhead can be compared in terms of trainable parameter count, total training time, and the number of training steps required to reach a prescribed validation-loss threshold.

Table~\ref{tab:model_size} summarizes the trainable parameter counts of the three architectures. The pure PINN has the smallest model size, while the LSTM-PINN introduces additional recurrent parameters. The proposed RA-PINN has the largest parameter count because each residual-attention block contains both a residual-feature branch and an attention-gate branch, together with layer normalization.

\begin{table}[htbp]
\centering
\caption{Trainable parameter counts of different architectures.}
\label{tab:model_size}
\begin{tabular}{lc}
\hline
Model & Trainable parameters \\
\hline
Pure PINN & 83,589 \\
LSTM-PINN & 281,733 \\
RA-PINN & 415,365 \\
\hline
\end{tabular}
\end{table}

Table~\ref{tab:cost_comparison} reports the total training time and the number of training steps required to reach a validation loss below $10^{-6}$ for the three benchmark cases. As expected, RA-PINN requires a higher computational cost than the pure PINN and LSTM-PINN in terms of parameter count and wall-clock training time. However, it consistently achieves the best final accuracy and better recovery of difficult local structures. These results indicate that the proposed gated residual-attention design improves solution quality at the price of additional computational overhead, providing a favorable accuracy--cost trade-off for problems containing irregular interfaces, localized high-gradient regions, and multi-peak spatial structures.

\begin{table}[htbp]
\centering
\caption{Computational cost comparison under the same training budget of 50,000 steps. The convergence speed is measured by the number of training steps required to reach a validation loss below $10^{-6}$.}
\label{tab:cost_comparison}
\begin{tabular}{lccc}
\hline
Case / Model & Training time (s) & Time per 1000 steps (s) & Steps to val.\ loss $<10^{-6}$ \\
\hline
Case 1 / Pure PINN & 17956.17 & 359.12 & 35200 \\
Case 1 / LSTM-PINN & 32335.94 & 646.72 & 23400 \\
Case 1 / RA-PINN & 61072.60 & 1221.45 & 25800 \\
\hline
Case 2 / Pure PINN & 13277.31 & 265.55 & 36800 \\
Case 2 / LSTM-PINN & 21951.32 & 439.03 & 25000 \\
Case 2 / RA-PINN & 61150.25 & 1223.00 & 31600 \\
\hline
Case 3 / Pure PINN & 10857.27 & 217.15 & 29800 \\
Case 3 / LSTM-PINN & 16918.38 & 338.37 & 16600 \\
Case 3 / RA-PINN & 49637.31 & 992.75 & 24800 \\
\hline
\end{tabular}
\end{table}

The results show that RA-PINN is the most computationally expensive architecture among the three compared models. Nevertheless, its superior prediction accuracy and improved recovery of difficult local structures indicate that the added model complexity is justified for the present benchmark problems. Compared with the pure PINN, RA-PINN consistently provides substantially better final solution quality, although at the cost of longer wall-clock training time. The LSTM-PINN offers an intermediate trade-off between complexity and efficiency, but its final accuracy remains lower than that of RA-PINN.

\subsection{Case 1: Oblique asymmetric interface}

Case 1 is designed to evaluate the ability of the models to reconstruct an oblique asymmetric transition interface. For this type of benchmark, the main difficulty does not lie in recovering a globally smooth field, but in preserving the orientation, sharpness, and local geometry of the inclined transition structure.

The pure PINN can recover the general field distribution reasonably well, but its predicted interface tends to be smoother than the exact one, and local directional transition details are partially weakened. This indicates that the plain MLP-based structure is able to capture the large-scale spatial pattern but still has difficulty preserving sharp directional variations.

The LSTM-PINN improves the prediction quality compared with the pure PINN. The global field structure becomes closer to the reference solution, and the interface-related regions are reconstructed more accurately. However, the predicted transition layer may still exhibit local smoothing or slight displacement in regions where the oblique interface changes rapidly.

In contrast, RA-PINN provides the closest agreement with the exact solution in Case 1. The predicted field better preserves the inclination of the transition layer, and the local interface geometry remains sharper and more consistent with the reference field. The corresponding absolute error distribution is also more localized, indicating that the residual-attention structure can more effectively focus on difficult directional transition regions while maintaining global field consistency.

\begin{table}[tbp]
\centering
\caption{Detailed field-wise error metrics of pure PINN, LSTM-PINN, and RA-PINN for Case 1.}
\label{tab:case1_full_metrics}
\resizebox{\linewidth}{!}{%
\begin{tabular}{llccccc}
\toprule
Model & Field & RMSE & MSE & MAE & L2 error & MaxAbs \\
\midrule
pure PINN & $u$   & $2.4970787653\times10^{-4}$ & $6.2354023601\times10^{-8}$ & $2.0456853753\times10^{-4}$ & $4.7697818942\times10^{-4}$ & $1.1574217905\times10^{-3}$ \\
pure PINN & $v$   & $2.4256825783\times10^{-4}$ & $5.8839359706\times10^{-8}$ & $1.7469541166\times10^{-4}$ & $4.5345715488\times10^{-4}$ & $1.7968652120\times10^{-3}$ \\
pure PINN & $p$   & $2.2956283192\times10^{-4}$ & $5.2699093797\times10^{-8}$ & $1.7170985601\times10^{-4}$ & $3.2727606321\times10^{-4}$ & $1.3568949394\times10^{-3}$ \\
pure PINN & $T$   & $2.7437436304\times10^{-4}$ & $7.5281291093\times10^{-8}$ & $2.1719082571\times10^{-4}$ & $4.0332392258\times10^{-4}$ & $2.2825462525\times10^{-3}$ \\
pure PINN & $\phi$ & $1.6601871396\times10^{-4}$ & $2.7562213385\times10^{-8}$ & $1.1575605415\times10^{-4}$ & $2.7383089881\times10^{-4}$ & $1.8501977411\times10^{-3}$ \\
\midrule
LSTM-PINN & $u$   & $1.5523526683\times10^{-4}$ & $2.4097988067\times10^{-8}$ & $1.2563385504\times10^{-4}$ & $2.9652182996\times10^{-4}$ & $7.4530149063\times10^{-4}$ \\
LSTM-PINN & $v$   & $1.6275392764\times10^{-4}$ & $2.6488840961\times10^{-8}$ & $1.2144789563\times10^{-4}$ & $3.0425222835\times10^{-4}$ & $1.5832172097\times10^{-3}$ \\
LSTM-PINN & $p$   & $1.9467247855\times10^{-4}$ & $3.7897373903\times10^{-8}$ & $1.4829348847\times10^{-4}$ & $2.7753465952\times10^{-4}$ & $1.3429940790\times10^{-3}$ \\
LSTM-PINN & $T$   & $2.0369428361\times10^{-4}$ & $4.1491361176\times10^{-8}$ & $1.5068139551\times10^{-4}$ & $2.9942585219\times10^{-4}$ & $1.2442648771\times10^{-3}$ \\
LSTM-PINN & $\phi$ & $1.2358006652\times10^{-4}$ & $1.5272032842\times10^{-8}$ & $7.4934252722\times10^{-5}$ & $2.0383268780\times10^{-4}$ & $1.5220560771\times10^{-3}$ \\
\midrule
RA-PINN & $u$   & $\mathbf{6.2835230328\times10^{-5}}$ & $\mathbf{3.9482661704\times10^{-9}}$ & $\mathbf{4.9700666003\times10^{-5}}$ & $\mathbf{1.2002438534\times10^{-4}}$ & $\mathbf{2.9846860760\times10^{-4}}$ \\
RA-PINN & $v$   & $\mathbf{6.2634344166\times10^{-5}}$ & $\mathbf{3.9230610691\times10^{-9}}$ & $\mathbf{5.1557853806\times10^{-5}}$ & $\mathbf{1.1708865685\times10^{-4}}$ & $\mathbf{4.1904946729\times10^{-4}}$ \\
RA-PINN & $p$   & $\mathbf{6.7924299816\times10^{-5}}$ & $\mathbf{4.6137105055\times10^{-9}}$ & $\mathbf{5.0777947496\times10^{-5}}$ & $\mathbf{9.6836222373\times10^{-5}}$ & $\mathbf{3.3560323648\times10^{-4}}$ \\
RA-PINN & $T$   & $\mathbf{6.2120216601\times10^{-5}}$ & $\mathbf{3.8589213106\times10^{-9}}$ & $\mathbf{5.0058036628\times10^{-5}}$ & $\mathbf{9.1315271418\times10^{-5}}$ & $\mathbf{3.5712526682\times10^{-4}}$ \\
RA-PINN & $\phi$ & $\mathbf{3.6211471425\times10^{-5}}$ & $\mathbf{1.3112706628\times10^{-9}}$ & $\mathbf{2.5909125838\times10^{-5}}$ & $\mathbf{5.9727120703\times10^{-5}}$ & $\mathbf{3.0511187066\times10^{-4}}$ \\
\bottomrule
\end{tabular}%
}
\end{table}

Table~\ref{tab:case1_full_metrics} confirms that RA-PINN achieves the lowest error for all five physical variables in Case 1.

\begin{figure}[p]
\centering
\includegraphics[width=\textwidth,height=0.88\textheight,keepaspectratio]{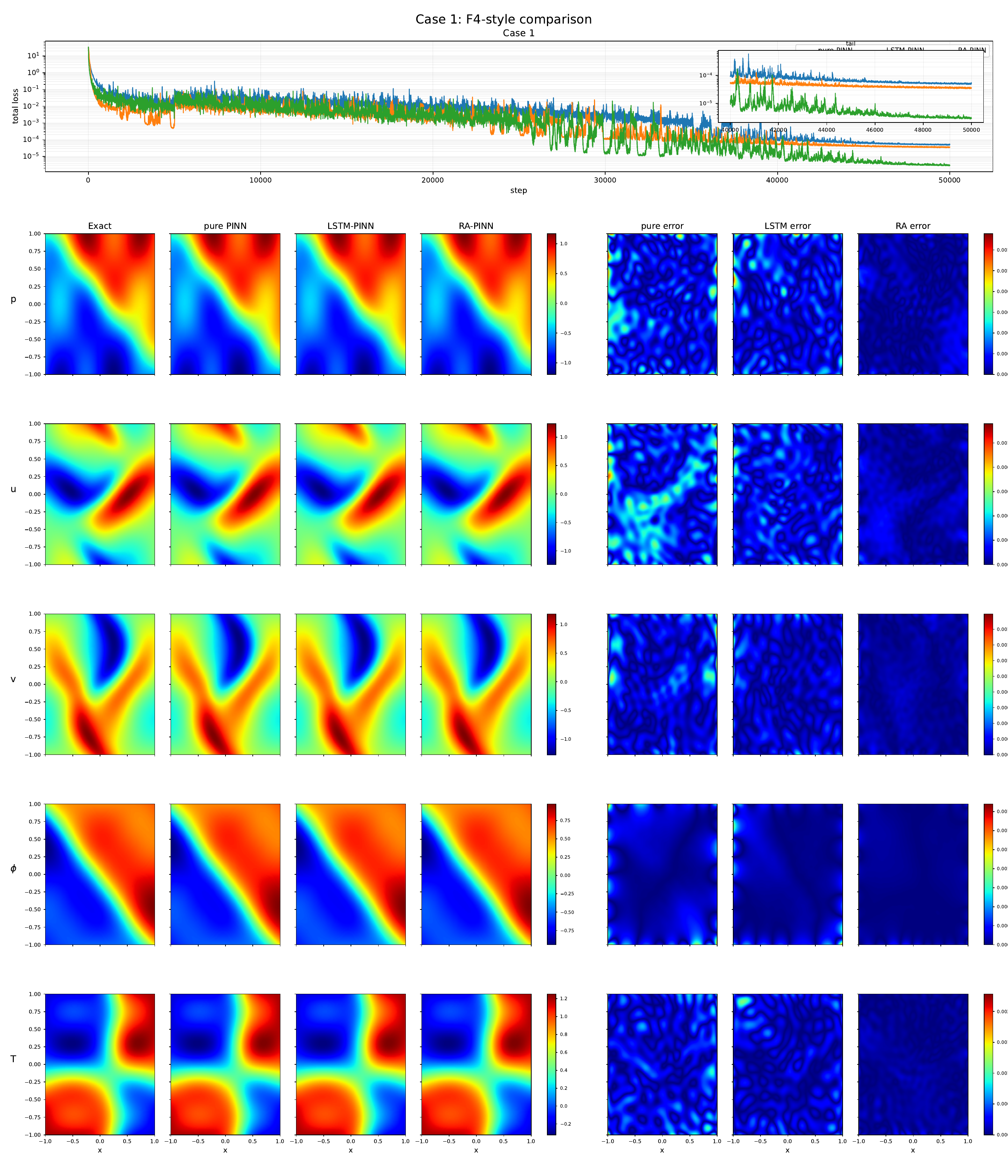}
\caption{Visual comparison for Case 1, including field prediction, absolute error, and log-scale loss evolution.}
\label{fig:case1_f4}
\end{figure}

Fig.\ref{fig:case1_f4} provides a visual comparison for Case 1. The proposed model yields the closest field reconstruction to the exact solution and the most localized error distribution across all five physical variables. The log-scale loss curve at the top of the figure further supports its more favorable convergence behavior.

\FloatBarrier

These observations suggest that the residual-attention mechanism is particularly effective for oblique and non-axisymmetric interface reconstruction, which is highly relevant to problems involving irregular transition boundaries or directional local discontinuities.

\subsection{Case 2: Bipolar high-gradient charge layer}

Case 2 focuses on a bipolar high-gradient charge layer, which is characterized by rapid positive-negative variation in a narrow region. For this benchmark, the key challenge is whether the model can accurately capture steep local gradients without sacrificing overall stability.

For the pure PINN, the predicted field captures the global trend of the solution, but the narrow high-gradient layer is not fully resolved. The local transition tends to be smoothed, and the error is mainly concentrated around the region where the exact field changes most rapidly. This behavior is typical for standard PINNs when sharp local structures occupy only a small portion of the domain.

The LSTM-PINN performs better than the pure PINN in Case 2. It improves the reconstruction of the local charge-transition layer and reduces the error around the steep-gradient region. This indicates that the LSTM-based hidden-state propagation is beneficial for representing complex local variation to some extent.

Nevertheless, RA-PINN still achieves the best performance among the three models. Its predicted field reproduces the narrow bipolar transition zone more accurately, and the corresponding high-gradient structure is preserved with less distortion. In addition, the local error near the steep transition region is reduced more significantly than in the other two models. This result shows that the attention-enhanced representation helps the network place more emphasis on the critical local layer, while the residual structure helps maintain stable global approximation.

\begin{table}[tbp]
\centering
\caption{Detailed field-wise error metrics of pure PINN, LSTM-PINN, and RA-PINN for Case 2.}
\label{tab:case2_full_metrics}
\resizebox{\linewidth}{!}{%
\begin{tabular}{llccccc}
\toprule
Model & Field & RMSE & MSE & MAE & L2 error & MaxAbs \\
\midrule
pure PINN & $u$   & $3.0156560659\times10^{-4}$ & $9.0941815076\times10^{-8}$ & $2.3799995044\times10^{-4}$ & $6.2461528455\times10^{-4}$ & $1.7785283887\times10^{-3}$ \\
pure PINN & $v$   & $3.7036671303\times10^{-4}$ & $1.3717150212\times10^{-7}$ & $2.7956355768\times10^{-4}$ & $1.1471127592\times10^{-3}$ & $1.5457147247\times10^{-3}$ \\
pure PINN & $p$   & $3.0276080580\times10^{-4}$ & $9.1664105531\times10^{-8}$ & $2.3874769851\times10^{-4}$ & $7.7773919119\times10^{-4}$ & $2.1434179455\times10^{-3}$ \\
pure PINN & $T$   & $5.0415074756\times10^{-4}$ & $2.5416797627\times10^{-7}$ & $3.8864828392\times10^{-4}$ & $1.0343430521\times10^{-3}$ & $3.2023466574\times10^{-3}$ \\
pure PINN & $\phi$ & $4.4384581524\times10^{-4}$ & $1.9699910771\times10^{-7}$ & $2.8529446367\times10^{-4}$ & $1.6084356133\times10^{-3}$ & $3.5818804588\times10^{-3}$ \\
\midrule
LSTM-PINN & $u$   & $2.0322706748\times10^{-4}$ & $4.1301240955\times10^{-8}$ & $1.5613305732\times10^{-4}$ & $4.2093239351\times10^{-4}$ & $1.1769943596\times10^{-3}$ \\
LSTM-PINN & $v$   & $1.8236400841\times10^{-4}$ & $3.3256631563\times10^{-8}$ & $1.4371600483\times10^{-4}$ & $5.6482419586\times10^{-4}$ & $1.1044373247\times10^{-3}$ \\
LSTM-PINN & $p$   & $1.8243369039\times10^{-4}$ & $3.3282051391\times10^{-8}$ & $1.4044475794\times10^{-4}$ & $4.6864002240\times10^{-4}$ & $1.5380125267\times10^{-3}$ \\
LSTM-PINN & $T$   & $2.7094912710\times10^{-4}$ & $7.3413429476\times10^{-8}$ & $2.0327172901\times10^{-4}$ & $5.5589394331\times10^{-4}$ & $1.2541266343\times10^{-3}$ \\
LSTM-PINN & $\phi$ & $2.2134720652\times10^{-4}$ & $4.8994585833\times10^{-8}$ & $1.4270375884\times10^{-4}$ & $8.0213154577\times10^{-4}$ & $1.8010493812\times10^{-3}$ \\
\midrule
RA-PINN & $u$   & $\mathbf{9.0938747082\times10^{-5}}$ & $\mathbf{8.2698557209\times10^{-9}}$ & $\mathbf{7.2661350012\times10^{-5}}$ & $\mathbf{1.8835613261\times10^{-4}}$ & $\mathbf{5.8873124306\times10^{-4}}$ \\
RA-PINN & $v$   & $\mathbf{6.7294925175\times10^{-5}}$ & $\mathbf{4.5286069543\times10^{-9}}$ & $\mathbf{5.2321398324\times10^{-5}}$ & $\mathbf{2.0842819989\times10^{-4}}$ & $\mathbf{3.2947750881\times10^{-4}}$ \\
RA-PINN & $p$   & $\mathbf{7.0506451761\times10^{-5}}$ & $\mathbf{4.9711597400\times10^{-9}}$ & $\mathbf{5.2951017369\times10^{-5}}$ & $\mathbf{1.8111865775\times10^{-4}}$ & $\mathbf{3.8034931781\times10^{-4}}$ \\
RA-PINN & $T$   & $\mathbf{1.0443281494\times10^{-4}}$ & $\mathbf{1.0906212836\times10^{-8}}$ & $\mathbf{7.9756279593\times10^{-5}}$ & $\mathbf{2.1426003445\times10^{-4}}$ & $\mathbf{8.7595184220\times10^{-4}}$ \\
RA-PINN & $\phi$ & $\mathbf{6.7110974819\times10^{-5}}$ & $\mathbf{4.5038829411\times10^{-9}}$ & $\mathbf{4.9306988338\times10^{-5}}$ & $\mathbf{2.4320085542\times10^{-4}}$ & $\mathbf{6.0324814150\times10^{-4}}$ \\
\bottomrule
\end{tabular}%
}
\end{table}

Table~\ref{tab:case2_full_metrics} confirms that RA-PINN achieves the lowest error for all five physical variables in Case 2.

\begin{figure}[p]
\centering
\includegraphics[width=\textwidth,height=0.88\textheight,keepaspectratio]{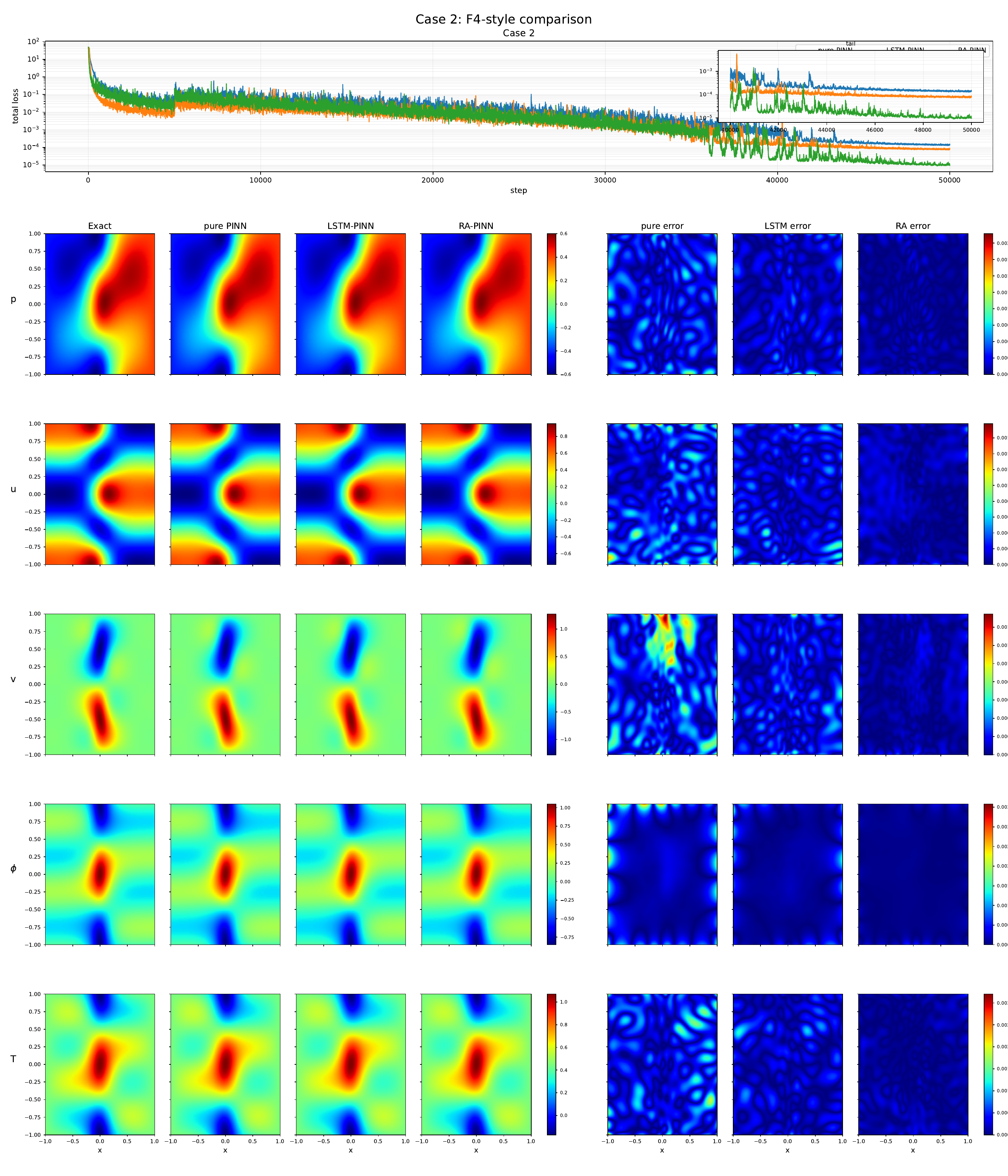}
\caption{Visual comparison for Case 2, including field prediction, absolute error, and log-scale loss evolution.}
\label{fig:case2_f4}
\end{figure}

Fig.\ref{fig:case2_f4} shows the visual comparison for Case 2. The proposed RA-PINN preserves the narrow bipolar high-gradient layer more accurately across the coupled fields and produces smaller error concentrations around the steep transition region. The corresponding log-scale loss curve also indicates more favorable convergence behavior.

\FloatBarrier

From an application point of view, this advantage is important because narrow high-gradient layers are often the most sensitive part of a coupled-field system. The ability of RA-PINN to better recover these localized regions suggests that it may provide more reliable predictions for interface-dominated or charge-layer-dominated problems.

\subsection{Case 3: Multi-peak Gaussian charge migration field}

Case 3 represents a multi-peak Gaussian charge migration field, which is intended to test the capability of the models in reconstructing several spatially separated hotspots under a highly nonuniform source distribution. Compared with Cases 1 and 2, this benchmark places more emphasis on the simultaneous preservation of multiple local extrema and complex migration paths.

The pure PINN can recover the broad structure of the field, but its performance becomes weaker when several peaks must be represented at the same time. In the predicted results, some hotspot regions are less distinct than in the reference solution, and the local intensity distribution may be underestimated. As a result, although the large-scale pattern is reproduced, the local multi-peak structure is not fully preserved.

The LSTM-PINN improves the reconstruction of multiple peaks compared with the pure PINN. The predicted field is more consistent with the exact field, and the main hotspot regions become more distinguishable. However, some local peak intensities and surrounding transition zones still show noticeable deviations.

Among the three models, RA-PINN provides the most accurate reconstruction of the multi-peak field. The predicted hotspots are more clearly separated, their locations are more consistent with the reference solution, and the surrounding local transition patterns are better preserved. The corresponding error maps show that the error concentration around peak regions is more limited than in the pure PINN and LSTM-PINN cases.

\begin{table}[tbp]
\centering
\caption{Detailed field-wise error metrics of pure PINN, LSTM-PINN, and RA-PINN for Case 3.}
\label{tab:case3_full_metrics}
\resizebox{\linewidth}{!}{%
\begin{tabular}{llccccc}
\toprule
Model & Field & RMSE & MSE & MAE & L2 error & MaxAbs \\
\midrule
pure PINN & $u$   & $1.4478986158\times10^{-4}$ & $2.0964104016\times10^{-8}$ & $1.1213482514\times10^{-4}$ & $9.0037555995\times10^{-4}$ & $1.1711853201\times10^{-3}$ \\
pure PINN & $v$   & $1.5814889550\times10^{-4}$ & $2.5011073148\times10^{-8}$ & $1.1966147912\times10^{-4}$ & $7.7980119679\times10^{-4}$ & $1.0195566536\times10^{-3}$ \\
pure PINN & $p$   & $1.7245253626\times10^{-4}$ & $2.9739877262\times10^{-8}$ & $1.3910822670\times10^{-4}$ & $1.0619735677\times10^{-3}$ & $9.2480388835\times10^{-4}$ \\
pure PINN & $T$   & $1.7610631261\times10^{-4}$ & $3.1013433340\times10^{-8}$ & $1.3982440790\times10^{-4}$ & $2.9373557793\times10^{-4}$ & $1.0144143251\times10^{-3}$ \\
pure PINN & $\phi$ & $2.0378705799\times10^{-4}$ & $4.1529165004\times10^{-8}$ & $1.4957254228\times10^{-4}$ & $3.8650885861\times10^{-4}$ & $1.4429612194\times10^{-3}$ \\
\midrule
LSTM-PINN & $u$   & $1.3841244092\times10^{-4}$ & $1.9158003801\times10^{-8}$ & $1.0809847333\times10^{-4}$ & $8.6071757812\times10^{-4}$ & $8.3531321207\times10^{-4}$ \\
LSTM-PINN & $v$   & $1.9267935745\times10^{-4}$ & $3.7125334789\times10^{-8}$ & $1.5151570978\times10^{-4}$ & $9.5006413458\times10^{-4}$ & $7.4113156340\times10^{-4}$ \\
LSTM-PINN & $p$   & $1.6793632258\times10^{-4}$ & $2.8202608441\times10^{-8}$ & $1.3376446069\times10^{-4}$ & $1.0341624397\times10^{-3}$ & $1.0023009677\times10^{-3}$ \\
LSTM-PINN & $T$   & $1.5147656457\times10^{-4}$ & $2.2945149613\times10^{-8}$ & $1.1904563242\times10^{-4}$ & $2.5265452202\times10^{-4}$ & $7.9411962278\times10^{-4}$ \\
LSTM-PINN & $\phi$ & $1.4416504432\times10^{-4}$ & $2.0783560004\times10^{-8}$ & $8.0621494393\times10^{-5}$ & $2.7342789715\times10^{-4}$ & $1.5137702094\times10^{-3}$ \\
\midrule
RA-PINN & $u$   & $\mathbf{9.6025292509\times10^{-5}}$ & $\mathbf{9.2208568015\times10^{-9}}$ & $\mathbf{8.1251342020\times10^{-5}}$ & $\mathbf{5.9713315262\times10^{-4}}$ & $\mathbf{4.7308012142\times10^{-4}}$ \\
RA-PINN & $v$   & $\mathbf{6.2788881018\times10^{-5}}$ & $\mathbf{3.9424435795\times10^{-9}}$ & $\mathbf{5.1860498181\times10^{-5}}$ & $\mathbf{3.0959966181\times10^{-4}}$ & $\mathbf{3.0372059220\times10^{-4}}$ \\
RA-PINN & $p$   & $\mathbf{6.4397256073\times10^{-5}}$ & $\mathbf{4.1470065897\times10^{-9}}$ & $\mathbf{5.2275832184\times10^{-5}}$ & $\mathbf{3.9656235429\times10^{-4}}$ & $\mathbf{3.1531490186\times10^{-4}}$ \\
RA-PINN & $T$   & $\mathbf{7.8786876641\times10^{-5}}$ & $\mathbf{6.2073719308\times10^{-9}}$ & $\mathbf{6.5079200887\times10^{-5}}$ & $\mathbf{1.3141214759\times10^{-4}}$ & $\mathbf{3.5613606464\times10^{-4}}$ \\
RA-PINN & $\phi$ & $\mathbf{4.4721695992\times10^{-5}}$ & $\mathbf{2.0000300924\times10^{-9}}$ & $\mathbf{2.6776121953\times10^{-5}}$ & $\mathbf{8.4820556532\times10^{-5}}$ & $\mathbf{8.4016955380\times10^{-4}}$ \\
\bottomrule
\end{tabular}%
}
\end{table}

Table~\ref{tab:case3_full_metrics} confirms that RA-PINN achieves the lowest error for all five physical variables in Case 3.

\begin{figure}[p]
\centering
\includegraphics[width=\textwidth,height=0.88\textheight,keepaspectratio]{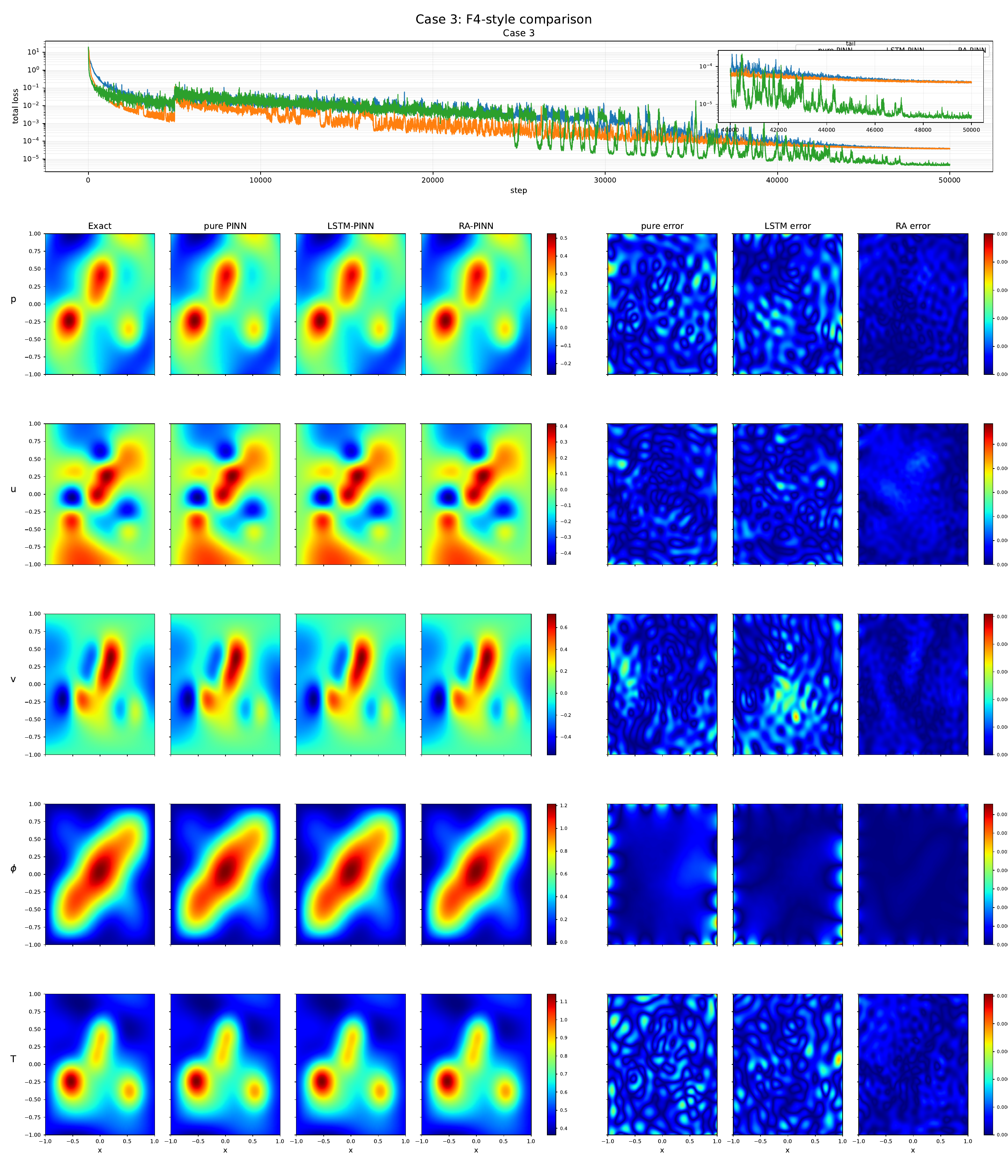}
\caption{Visual comparison for Case 3, including field prediction, absolute error, and log-scale loss evolution.}
\label{fig:case3_f4}
\end{figure}

Fig.\ref{fig:case3_f4} shows the visual comparison for Case 3. RA-PINN achieves the most accurate reconstruction of the multi-peak Gaussian migration field. The hotspot locations and intensities are better preserved across the five variables, while the associated absolute errors remain more localized. The log-scale loss evolution further supports the superior convergence of the proposed model.

\FloatBarrier

These findings indicate that the proposed residual-attention structure is highly beneficial for reconstructing spatially distributed complex local features. Since many practical coupled-field systems involve multiple interacting local maxima rather than a single transition structure, the superior performance of RA-PINN in Case 3 further supports its applicability to complex multi-physics field modeling.

\section{Engineering Implications}

The engineering implication of the present study is that reliable coupled-field prediction should be judged not only by global averaged error, but also by the accuracy with which locally critical structures are reconstructed. In many practical systems, the most important physical information is concentrated in irregular interfaces, localized steep-gradient layers, and multiple hotspot regions, because these structures are directly related to transport barriers, charge accumulation, thermal concentration, and local forcing intensity. Therefore, a model that appears globally accurate but fails in these critical local regions may still be insufficient for engineering use. From this perspective, the present results suggest that residual-attention enhancement provides a more suitable physics-informed modeling strategy for coupled multi-physics problems in which both global consistency and local structural fidelity are required.

The three benchmark cases further clarify this engineering meaning. Case 1 shows that oblique and asymmetric interface structures are difficult for conventional PINN architectures to recover accurately, even when the large-scale field trend is captured reasonably well. In practical engineering systems, interfaces are often geometrically irregular and directionally biased rather than simple, symmetric, or coordinate-aligned. The improved performance of RA-PINN in this case therefore suggests that the proposed framework is more suitable for interface-dominated coupled-field problems in which accurate interface position, directional transition sharpness, and local geometric consistency are important for physical interpretation.

Case 2 indicates that narrow high-gradient layers remain a major challenge for coupled-field prediction. In practical applications, such localized steep-gradient structures may correspond to thin transport layers, charge-separation zones, or confined regions with strong field variation. Although these regions occupy only a small portion of the spatial domain, they often dominate the sensitivity and reliability of the whole system response. The superior performance of RA-PINN in Case 2 therefore implies that the proposed architecture is particularly relevant for engineering scenarios in which reliable prediction of localized gradients is more important than merely reducing a global average error.

Case 3 further extends this implication to problems with multiple distributed hotspots and complex migration paths. In many realistic coupled systems, the target field contains several spatially separated local extrema rather than a single dominant transition structure. Such multi-peak patterns are challenging because the model must preserve several local responses simultaneously while maintaining a coherent global background. The better reconstruction achieved by RA-PINN in this case suggests that the proposed framework is especially suitable for engineering tasks involving distributed charge accumulation, multiple transport channels, or complex source interactions, where the relative location, intensity, and interaction of several local critical regions must be resolved in a stable manner.

Taken together, these results suggest that the advantage of RA-PINN is not limited to a single benchmark-specific pattern, but reflects a more general improvement in the capability of PINNs to model engineering-oriented coupled fields with complex local structures. Residual learning helps maintain stable large-scale feature propagation, while the gated attention mechanism improves sensitivity to spatially critical regions. Their combination provides a practical balance between global field consistency and local structural accuracy, which is particularly important in applications such as electro-thermal-fluid coupling, microscale transport regulation, interface-dominated transport processes, and multi-source field reconstruction. From an engineering viewpoint, the present study therefore supports residual-attention-enhanced PINNs as a promising framework for coupled multi-physics problems in which local structural reliability is of primary concern.

\section*{Conclusions}

In modern engineering applications such as electro-thermal coupled processes, the accurate prediction of complex multi-physics fields is frequently hindered by irregular interfaces, narrow high-gradient regions, and multi-peak transport structures. To overcome the limitations of standard deep learning models in resolving these critical local features, this study develops a novel intelligent surrogate modeling framework termed the Residual-Attention Physics-Informed Neural Network (RA-PINN). By seamlessly integrating residual learning for global field consistency with an attention-gate mechanism for localized feature sensitivity, this dual-branch architecture strictly maintains physics-constrained fidelity. The representational capability of the proposed method was rigorously validated against a standard pure PINN and an LSTM-PINN baseline through three representative benchmark cases encompassing asymmetric interfaces, bipolar charge layers, and multi-peak Gaussian migration fields. Numerical results unequivocally demonstrate that RA-PINN consistently achieves superior overall prediction performance and substantially higher accuracy in reconstructing complex, localized structural transitions without artificial smoothing. Consequently, future work will focus on integrating this powerful predictive architecture into real-time condition monitoring expert systems, where it holds tremendous potential to serve as the core intelligent inference engine within digital twin frameworks for dynamic real-world engineering systems.

\clearpage
\section*{Declaration of competing interest}
The authors declared that they have no conflicts of interest to this work. 

\section*{Acknowledgment}
This work is supported by the developing Project of Science and Technology of Jilin Province (20250102032JC).  

\section*{Data availability}
All the code for this article is available open access at a Github repository available at https://github.com/Uderwood-TZ/RA-PINN-for-Irregular-Interfaces-and-Multi-Peak-Transport-Fields.git.
\clearpage
\appendix
\section{Hyperparameter settings of the final residual-attention implementations}

This appendix summarizes the main hyperparameter settings used in the final residual-attention implementations for Cases 1, 2, and 3. These settings are extracted directly from the corresponding Python scripts and are provided here for reproducibility.

\subsection{Main training and architecture settings}

\begin{table}[h]
\centering
\caption{Main hyperparameter settings of the final residual-attention implementations for Cases 1, 2, and 3.}
\label{tab:ra_hyperparameters}
\resizebox{\linewidth}{!}{%
\begin{tabular}{lccc}
\toprule
Hyperparameter & Case 1 & Case 2 & Case 3 \\
\midrule
Random seed & 20260308 & 20260308 & 20260308 \\
Input dimension & 2 & 2 & 2 \\
Output dimension & 5 & 5 & 5 \\
Hidden dimension & 128 & 128 & 128 \\
Number of residual-attention blocks & 6 & 6 & 6 \\
MLP depth setting & 6 & 6 & 6 \\
Feature branch in each block & Linear--Tanh--Linear & Linear--Tanh--Linear & Linear--Tanh--Linear \\
Attention-gate branch in each block & Linear--Tanh--Linear--Sigmoid & Linear--Tanh--Linear--Sigmoid & Linear--Tanh--Linear--Sigmoid \\
Normalization & LayerNorm & LayerNorm & LayerNorm \\
Training steps & 50{,}000 & 50{,}000 & 50{,}000 \\
Optimizer & Adam & Adam & Adam \\
Initial learning rate & $8.0\times10^{-4}$ & $8.0\times10^{-4}$ & $8.0\times10^{-4}$ \\
Minimum learning rate & $1.0\times10^{-5}$ & $1.0\times10^{-5}$ & $1.0\times10^{-5}$ \\
Weight decay & $1.0\times10^{-10}$ & $1.0\times10^{-10}$ & $1.0\times10^{-10}$ \\
Maximum gradient norm & 1.0 & 1.0 & 1.0 \\
Learning-rate scheduler & CosineAnnealingLR & CosineAnnealingLR & CosineAnnealingLR \\
Scheduler $T_{\max}$ & 50{,}000 & 50{,}000 & 50{,}000 \\
Train ratio & 0.7 & 0.7 & 0.7 \\
Supervised grid size & $91\times91$ & $91\times91$ & $91\times91$ \\
Evaluation grid size & $181\times181$ & $181\times181$ & $181\times181$ \\
Number of collocation points & 16{,}000 & 16{,}000 & 16{,}000 \\
Boundary points per edge & 360 & 360 & 360 \\
Data batch size & 1024 & 1024 & 1024 \\
Collocation batch size & 2048 & 2048 & 2048 \\
Boundary batch size & 1024 & 1024 & 1024 \\
Logging interval & 500 & 500 & 500 \\
Validation interval & 200 & 200 & 200 \\
PDE loss weight & 1.0 & 1.0 & 1.0 \\
Data loss weight & 18.0 & 18.0 & 18.0 \\
Boundary loss weight & 25.0 & 25.0 & 25.0 \\
Warm-up PDE weight & 0.25 & 0.25 & 0.25 \\
Warm-up steps & 5{,}000 & 5{,}000 & 5{,}000 \\
Residual weight (continuity) & 2.0 & 2.0 & 2.0 \\
Residual weight (momentum-$x$) & 1.0 & 1.0 & 1.0 \\
Residual weight (momentum-$y$) & 1.0 & 1.0 & 1.0 \\
Residual weight (energy) & 1.0 & 1.0 & 1.0 \\
Residual weight ($\phi$) & 1.0 & 1.0 & 1.0 \\
Domain in $x$ & $[-1,1]$ & $[-1,1]$ & $[-1,1]$ \\
Domain in $y$ & $[-1,1]$ & $[-1,1]$ & $[-1,1]$ \\
Viscosity coefficient $\nu$ & 0.035 & 0.035 & 0.035 \\
Thermal diffusivity & 0.020 & 0.020 & 0.020 \\
Electric force coefficient & 0.45 & 0.45 & 0.45 \\
Thermal force coefficient & 0.22 & 0.22 & 0.22 \\
\bottomrule
\end{tabular}%
}
\end{table}
\clearpage
\bibliographystyle{model1-num-names}
\bibliography{main}

\end{document}